\documentclass[RNAAS]{aastex63}

\graphicspath{{./}{figures/}}
\usepackage[normalem]{ulem}
\begin{document}

\title{Resolving the Multiplicity of Exoplanet Host Stars in Gemini/NIRI Data}

\correspondingauthor{Kim Miskovetz}
\email{kmiskove@hawaii.edu}

\author{Kim Miskovetz}
\affiliation{Department of Physics and Astronomy, University of Hawai`i at Hilo, Hilo, HI 96720, USA}

\author[0000-0001-9823-1445]{Trent J.\ Dupuy}
\affiliation{Gemini Observatory, NSF's OIR Lab, Hilo, HI 96720, USA}
\affiliation{Institute for Astronomy, University of Edinburgh, Royal Observatory, Blackford Hill, Edinburgh, EH9 3HJ, UK}

\author[0000-0002-1043-8853]{Jessica Schonhut-Stasik}
\affiliation{Gemini Observatory, NSF's OIR Lab, Hilo, HI 96720, USA}
\affiliation{Department of Physics and Astronomy, Vanderbilt University, Nashville, TN 37235, USA}
\affiliation{FRIST Center for Autism, 2414 Highland Avenue, Suite 115, Nashville, TN 37212, USA}

\author[0000-0002-3481-9052]{Keivan G. Stassun}
\affiliation{Department of Physics and Astronomy, Vanderbilt University, Nashville, TN 37235, USA}

\keywords{Astrometric binary stars(79), Direct imaging(387), Planet hosting stars(1242), Visual binary stars(1777)}

\begin{abstract}
The majority of stars have one or more stellar companions. As exoplanets continue to be discovered, it is crucial to examine planetary systems to identify their stellar companions. By observing a change in proper motion, companions can be detected by the acceleration they induce on their host stars. We selected 701 stars from the Hipparcos--Gaia Catalog of Accelerations (HGCA) that have existing adaptive optics (AO) imaging data gathered with Gemini/NIRI. Of these, we examined 21 stars known to host planet candidates and reduced their archival NIRI data with Gemini's DRAGONS software. We assessed these systems for companions using the NIRI images as well as RUWE values in Gaia and accelerations in the HGCA. We detected three known visible companions and found two more systems with no visible companions but astrometric measurements indicating likely unresolved companions.
\end{abstract}

\section{} 

The majority of stars have one or more stellar companions \citep[e.g.,][]{2017Moe}, so our understanding of how planetary systems form and evolve is incomplete without accounting for multiple-star systems. Developing such an understanding is an ongoing challenge \citep[e.g.,][]{2019arXiv} that relies on comprehensive observational evidence over a wide range of separations, ranging from inside the scale of the solar-system \citep[$<$50\,{AU}; e.g.,][]{2016AJ} to much wider \citep[$\sim10^3$\,{AU}; e.g.,][]{2016ApJ}.

{\sl Gaia} observations present an opportunity to examine the multiplicity of planetary systems on both of these scales \citep{2018GaiaVizieR}.
Near or inside the angular resolution limit of {\sl Gaia} ($\approx$0\farcs1), poor astrometric fit as quantified by the Renormalised Unit Weight Error (RUWE) can indicate unresolved multiplicity \citep[e.g.,][]{2020arXiv}.
At wider separations, a discrepancy in proper motions at different epochs can be used to identify massive bodies by the acceleration they induce on the primary star \citep[e.g.,][]{2019A&A}.
Combined, these methods probe close-in multiples (1--10\,AU) and wide companions on long-period orbits ($\sim$30--$10^3$\,yr, corresponding to $\sim$10--100\,AU) for the many known planetary systems within $\sim$10--100\,pc. AO imaging can then be used to visually investigate wider companions that astrometry may not detect.

Gemini Observatory's Near InfraRed Imager (NIRI) is capable of imaging wavelengths from 
1.0--5.5\,\micron{} 
with a typical field of view of $20\arcsec\times20\arcsec$  
\citep{NIRI}.
Gemini Observatory's ALTtitude conjugate Adaptive optics for the InfraRed (ALTAIR; \citealp{2010Altair}) further improves the quality of these observations, allowing NIRI to resolve both close binaries down to the diffraction limit of the 8-m Gemini telescope, and wider binaries.

We used the Hipparcos--Gaia Catalog of Accelerations \citep[HGCA;][]{Brandt_2018} for target selection, as it contains a uniform analysis of astrometric acceleration for bright stars. The HGCA is a cross-calibration of the \textsl{Hipparcos} \citep{Perryman1997} and \textsl{Gaia} missions \citep{Gaia2016a,Gaia2016b,Gaia2016,Gaia2018}.

We performed an automated search of the Gemini Observatory Archive \citep[GOA;][]{GOA} in a $10\arcsec$ radius around the Equinox J2000 Epoch 2000.0 right ascension and declination values of all stars in the HGCA catalog, and selected those that had NIRI data. We cross-matched a list of the resulting 701 stars to the list of known or suspected exoplanets from \url{exoplanets.org} \citep{2014exoplanets}, resulting in 23 stars of interest, five of which are known multiple systems. HD~19994 is a triple system with a circumprimary exoplanet and a binary pair orbiting the primary at $\approx$2\arcsec \citep{2016wiegert}. KELT-2 and WASP-33 also have visual companions at $\approx$2\arcsec \citep{Beatty2012,2019Mugrauer}. HD~89744 has an L-dwarf companion $63.1\arcsec$ away \citep{2001Wilson}, and HD~125612 has a stellar companion $\sim1.5\arcmin$ away \citep{2009Mugrauer}, both well outside NIRI's field-of-view.

We used Gemini Observatory's data reduction software DRAGONS (Data Reduction for Astronomy from Gemini Observatory North and South) to reduce imaging data from NIRI \citep{2018DRAGONS.soft11002A}. Data we used was taken in one of five different filters: $K_{\rm cont}$ ($\lambda_c = 2.097\,\micron$; FWHM$ = 0.031\,\micron$), $H$ ($\lambda_c = 1.649\,\micron$; FWHM$ = 0.291\,\micron$), $CH_4s$ ($\lambda_c = 1.595\,\micron$; FWHM$ = 0.115\,\micron$), Br$\gamma$ ($\lambda_c = 2.168\,\micron$; FWHM$ = 0.034\,\micron$), and $H_{\rm cont}$ ($\lambda_c = 1.576\,\micron$; FWHM$ = 0.022\,\micron$) \citep{FilterP}. 
We reduced 21 of 23 stars of interest and visually examined them for companions. $\varepsilon$~Tau and HD~218396 could not be reduced because $\varepsilon$~Tau had insufficient frames, and HD~218396 was highly saturated. 

We recovered all three companions expected to be in the NIRI data (WASP-33, KELT-2, and HD~19994). 
We found background stars that were not co-moving with HD~17156, HD~168443, and HAT-P-11. Four targets had data taken in angular differential imaging mode, such that DRAGONS failed to properly rotate the images while stacking, resulting in primary stars with well-defined diffraction spikes but any companions smeared by sky rotation. For GJ~436, HD~130322, and GJ~876, we found no such smeared companions. For HD~19994, we estimated an approximate separation of $2.3\pm0.1\arcsec$ from the curved smear left by its known companion.

For systems with visible companions, we computed their angular separations and position angles. These data were taken with NIRI's f/32 camera with a pixel scale of $21.9\, \mathrm{mas\,pixel}^{-1}$. We assumed a global uncertainty of $0.05\, \mathrm{mas\,pixel}^{-1}$ on the pixel scale and $0.1^{\circ}$ on the orientation of NIRI \citep{Beck_2004}. For KELT-2, we used NIRI data from 2013~Apr~9~UT to measure an angular separation of $2.433 \pm 0.006\arcsec$ and a position angle of $332.7\pm0.1^{\circ}$, where the calibration uncertainties dominate over relative pixel position uncertainties.
For comparison, \cite{Beatty2012} found an angular separation of $2.29 \pm 0.05\arcsec$ and a position angle of $328.6 \pm 0.9^{\circ}$ on 2012~Apr~10~UT. The disagreement between these measurements is larger than expected for orbital motion of a $\approx$300-AU binary, indicating that the measurement uncertainties are likely somewhat underestimated. Using \textsl{Gaia} data from 2015, we computed an angular separation of $2.3792\arcsec$ and a position angle of $332.140^{\circ}$, which is between our measurement and that of \cite{Beatty2012}.
We also found an angular separation of $1.980 \pm 0.005\arcsec$ and a position angle of $277.0\pm0.1^{\circ}$ for WASP-33. \cite{Ngo2016} found a position angle of $276.4 \pm 0.2^{\circ}$ on 2010 Nov 29, a position angle of $276.247\pm 0.045^{\circ}$ on 2013 Aug 19, and an angular separation of $1.940 \pm 0.002$. 

Additional criteria that may indicate a system as a potential binary include HGCA $\chi^2$ values above 11.8 ($>3\sigma$) or RUWE values above 1.2 \citep{2021Stassun}. By combining information on these systems' HGCA $\chi^2$ values, \textsl{Gaia} RUWE values, and Gemini/NIRI imaging, we can obtain a nearly complete view of these systems. Our sample included only one system with high RUWE, $\iota$~Dra. This star is very bright ($G = 2.8471 \pm 0.0037$\,mag; \citealp{2018GaiaVizieR}), so its high RUWE is most likely due to saturation in {\sl Gaia}. Our sample also included two systems with significantly high HGCA $\chi^2$ values: 14~Her ($\chi^2 = 439.78$) and HD~168443 ($\chi^2 = 34.02$). Though the NIRI images for these systems show no companions, these $\chi^2$ values suggest that they may be worth investigating with higher-contrast data. For 14~Her, \cite{2007Wittenmyer} found evidence of a long-period outer companion, which likely explains its very high $\chi^2$. Likewise, the massive outer brown dwarf in the HD~168443 system \citep{2011Pilyavsky} may explain its high $\chi^2$.

\bibliographystyle{aasjournal.bst}
\bibliography{bib.bib}
\clearpage

\begin{turnpage}
\begin{deluxetable}{lcccccccccclll}




\tablecaption{Gemini NIRI Observations of Planet Hosts}

\tablenum{1}

\tablehead{\colhead{Name} & \colhead{HIP ID} & \colhead{Method} & \colhead{HGCA $\chi^2$} & \colhead{RUWE} & \colhead{$d$} & \colhead{Date} & \colhead{Filter} & \colhead{$N_{\rm exp} \times t_{\rm exp}$} & \colhead{$K_S$} & \colhead{$N_{\rm comp}$} & \colhead{Program ID} & \colhead{PI} \\ 
\colhead{} & \colhead{} & \colhead{} & \colhead{} & \colhead{} & \colhead{(pc)} & \colhead{(UT)} & \colhead{} & \colhead{(s)} & \colhead{(mag)} & \colhead{} & \colhead{} & \colhead{} } 

\startdata
WASP-33 & 11397 & Transit & 0.48 & 0.88 & $121.9\pm1.0$\phn & 08/29/2012 & Kcon(209)\_G0217 & 9x10.4 & $7.468\pm0.024$ & 1 & GN-2012B-Q-118 & Quinn\\
HD 17156 & 13192 & RV & 1.38 & 1.00 & $78.09\pm0.24$ & 09/05/2012 & Kcon(209)\_G0217 & 9x5.6 & $6.763\pm0.024$ & 0 & GN-2012B-Q-118 & Quinn \\
HD 19994 & 14954 & RV & 1.67 & 1.13 & $22.52\pm0.10$ & 08/31/2005 & PK50\_G0201 & 148x30 & $3.75\pm0.24$ & 1 & GN-2005B-Q-4 & Doyon \\
$\varepsilon$ Eri & 16537 & RV & 1.56 & 0.92 & \phn$3.203\pm0.005$ & 09/08/2005 & PK50\_G0201 & 20x30 & $1.601\pm0.060$ & 0 & GN-2005B-Q-4 & Doyon \\
HD 34445 & 24681 & RV & 0.43 & 1.05 & $46.09\pm0.10$ & 02/23/2010 & H\_G0203 & 16x0.5 & $5.790\pm0.023$ & 0 & GN-2010A-SV-101 & Engineering \\
KELT-2 & 29301 & Transit & 1.67 & 0.96 & $134.06\pm0.80$\phn & 04/09/2013 & Kcon(209)\_G0217 & 9x7.2 & $7.346\pm0.031$ & 1 & GN-2013A-Q-54 & Quinn \\
HD 50499 & 32970 & RV & 0.95 & 0.98 & $46.28\pm0.06$ & 03/06/2008 & CH4(short) & 6x25 & $5.836\pm0.016$ & 0 & GN-2008A-Q-95 & Croll \\
HD 89744 & 50786 & RV & 1.54 & 0.65 & $38.64\pm0.11$ & 06/23/2005 & Brgamma\_G0218 & 9x10 & $4.454\pm0.021$ & 0 & GN-2005A-Q-22 & Thomas \\
GJ 433 & 56528 & RV & 6.20 & 1.11 & \phn$9.065\pm0.004$ & 05/10/2009 & PK50\_G0201 & 4x50 & $5.623\pm0.021$ & 0 & GN-2009A-Q-94 & Dieterich \\
GJ 436 & 57087 & RV & 2.00 & 1.18 & \phn$9.75\pm0.01$ & 03/23/2008 & CH4(short) & 11x25 & $6.073\pm0.016$ & 0 & GN-2008A-Q-95 & Croll \\
HD 102195 & 57370 & RV & 0.25 & 0.82 & $29.34\pm0.05$ & 03/18/2006 & CH4(short) & 90x30 & $6.151\pm0.018$ & 0 & GN-2006A-Q-5 & Doyon \\
HD 125612 & 70123 & RV & 1.39 & 1.02 & $57.62\pm0.16$ & 03/23/2008 & PK50\_G0201 & 5x25 & $6.838\pm0.026$ & 0 & GN-2008A-Q-95 & Croll \\
HD 130322 & 72339 & RV & 2.37 & 1.03 & $31.88\pm0.07$ & 05/15/2006 & PK50\_G0201 & 10x30 & $6.234\pm0.023$ & 0 & GN-2006A-Q-5 & Doyon \\
HD 134987 & 74500 & RV & 1.40 & 0.89 & $26.18\pm0.05$ & 03/24/2008 & PK50\_G0201 & 4x25 & $4.882\pm0.016$ & 0 & GN-2008A-Q-95 & Croll \\
$\iota$ Dra & 75458 & RV & 1.09 & 1.46 & $31.65\pm0.30$ & 03/22/2008 & H-con(157) & 6x25 & $0.701\pm0.041$ & 0 & GN-2008A-Q-95 & Croll \\
14 Her & 79248 & RV & 439.78 & 0.98 & $17.93\pm0.01$ & 03/24/2008 & CH4(short) & 11x25 & $4.714\pm0.016$ & 0 & GN-2008A-Q-95 & Croll \\
HAT-P-2 & 80076 & Transit & 1.70 & 1.01 & $127.77\pm0.40$\phn & 07/12/2012 & Kcon(209) & 4x11.22 & $7.603\pm0.020$ & 0 & GN-2012A-Q-99 & Quinn \\
HD 168443 & 89844 & RV & 34.02 & 0.94 & $39.62\pm0.12$ & 02/26/2012 & H\_G0203 & 10x60 & $5.211\pm0.015$ & 0\tablenotemark{*} & GN-2012A-Q-51 & Wahhaj \\
HAT-P-11 & 97657 & Transit & 9.61 & 0.98 & $37.76\pm0.03$ & 07/07/2012 & Kcon(209) & 4x6.15 & $7.009\pm0.020$ & 0\tablenotemark{*} & GN-2012A-Q-99 & Quinn \\
GJ 849 & 109388 & RV & 0.74 & 0.85 & \phn$8.800\pm0.004$ & 06/19/2008 & CH4(short) & 11x25 & $5.594\pm0.017$ & 0 & GN-2008A-Q-95 & Croll \\
GJ 876 & 113020 & RV & 2.01 & 1.13 & \phn$4.675\pm0.002$ & 08/21/2005 & CH4(short) & 90x30 & $5.010\pm0.021$ & 0 & GN-2005B-Q-4 & Doyon  
\enddata


\tablenotetext{*}{There are no physically associated companions, but there are other stars in the NIRI field-of-view.}

\tablecomments{The following programs have had their data published previously: GN-2005B-Q-4 and GN-2006A-Q-5 \citep{Galicher2016}, and GN-2009A-Q-94 \citep{Dieterich2013}.} 

\tablerefs{All $K_S$ values are from \cite{2003yCat}, with the exception of those of $\varepsilon$~Eri and $\iota$~Dra, which are from \cite{Adams2017}.}

\end{deluxetable}
 \label{table1}
\end{turnpage}
\clearpage
\global\pdfpageattr\expandafter{\the\pdfpageattr/Rotate 90}
\end{document}